\documentclass[aps,prl,twocolumn,superscriptaddress,showpacs,amsmath,amssymb,floatfix]{revtex4}

\usepackage{graphicx}
\usepackage{dcolumn}
\usepackage{bm}

\begin{document}


\title{Quantum Phase Transition in the Itinerant 
Antiferromagnet (V$_{0.9}$Ti$_{0.1}$)$_2$O$_3$}

\author{Hiroaki Kadowaki}
\email{kadowaki@comp.metro-u.ac.jp}
\affiliation{Department of Physics, 
Tokyo Metropolitan University, 
Hachioji, Tokyo 192-0397, Japan}

\author{Kiyoichiro Motoya}
\affiliation{Department of Physics, 
Tokyo University of Science, Noda, Chiba 278-8510, Japan}

\author{Taku J. Sato}
\affiliation{NSL, Institute for Solid State Physics, 
University of Tokyo, 
Tokai, Ibaraki 319-1106, Japan}

\author{J. W. Lynn}
\affiliation{NIST Center for Neutron Research, 
Gaithersburg, Maryland 20899-6102, USA}

\author{J. A. Fernandez-Baca}
\affiliation{Neutron Scattering Science Division, 
Oak Ridge National Laboratory, 
Oak Ridge, Tennessee 37831-6393, USA}

\author{Jun Kikuchi}
\affiliation{Department of Physics, 
Meiji University, Kawasaki, Kanagawa 214-8571, Japan} 

\date{\today}

\begin{abstract}
%
%
Quantum-critical behavior of the itinerant electron 
antiferromagnet (V$_{0.9}$Ti$_{0.1}$)$_2$O$_3$ 
has been studied by single-crystal neutron scattering. 
By directly observing antiferromagnetic spin fluctuations 
in the paramagnetic phase, 
we have shown that the characteristic energy 
depends on temperature as $c_1 + c_2 T^{3/2}$, 
where $c_1$ and $c_2$ are constants. 
This $T^{3/2}$ dependence demonstrates that 
the present strongly correlated $d$-electron 
antiferromagnet clearly shows the criticality of 
the spin-density-wave quantum phase transition 
in three space dimensions. 
\end{abstract}



\maketitle

In recent years, novel viewpoints of matter have been 
exploited by quantum phase transitions (QPT)
\cite{SachdevBook99,Laughlin01}, 
zero-temperature 
second-order 
phase transitions 
tuned by pressure or other controlling parameters. 
Around a QPT, the state of matter is characterized by 
singular behavior of fluctuating order parameters 
having both quantum mechanical and thermal origins. 
Quantum phase transitions are investigated in broad fields ranging 
from high temperature superconductors \cite{Moriya-Ueda03,Sachdev03},
metal-insulator transitions \cite{Imada98} to 
heavy fermions \cite{Stewart01,Lohneysen07}. 
Although a number of QPTs have been investigated 
experimentally and theoretically, many problems are 
under controversial debates. 

A QPT separating a ferromagnetic 
(FM) 
or 
antiferromagnetic (AFM) state to a paramagnetic state in an itinerant 
electron system has been studied for decades. 
Its theory was first developed by Moriya and 
coworkers \cite{MoriyaBook85,Moriya73,Hasegawa74}. 
The modern formulation of this theory using renormalization group 
techniques was provided by Hertz \cite{Hertz76,Millis93}. 
The theoretical predictions of the 
FM 
QPT are in general 
supported by the experimental studies of, for instance, 
$d$-electron FM metals 
MnSi and 
ZrZn$_2$ \cite{MoriyaBook85,Ishikawa85,Lonzarich84}. 
However recent studies of 
the FM 
QPT have shown 
that there are important perturbative effects 
closer to the critical point 
\cite{Pfleiderer01,Lohneysen07,Belitz05}. 

For the 
itinerant 
AFM QPT, referred to as the spin density wave (SDW) QPT, 
the problem is more complicated and is not settled. 
Experimentally, thermodynamic and transport properties 
studied on, e.g., 
$d$-electron AFM metals
$\beta$-Mn, V$_3$Se$_4$ \cite{MoriyaBook85} and 
$f$-electron 
AFM 
heavy fermions \cite{Stewart01,Lohneysen07} 
are in rough agreement with theories of the SDW QPT. 
However most neutron scattering studies 
seem to contradict expectations of the SDW QPT \cite{SachdevBook99}. 
For example, observed AFM spin fluctuations of 
the heavy fermion CeCu$_{6-x}$Au$_{x}$ \cite{Schroder00} 
exhibit $E/T$ scaling, 
suggesting the existence of a new type of QPT 
\cite{SachdevBook99,Lohneysen07,Si01,Coleman07}. 
On the other hand, our recent neutron scattering study on 
the heavy fermion 
Ce(Ru$_{1-x}$Rh$_x$)$_2$Si$_{2}$ is consistent 
with the SDW QPT with no indication of $E/T$ scaling \cite{Kado06}. 
Therefore there are many open questions on QPTs for itinerant 
antiferromagnets, such as, 
whether the SDW QPT can be applicable to the itinerant 
$d$- and $f$-electron AFM systems, 
or how fundamentally new QPTs are formulated to account for 
the complexity of experimental data 
of these itinerant systems 
\cite{Stewart01,Lohneysen07,Belitz05,Coleman07,Gegenwart07,Senthil04}.

\begin{figure}
\begin{center}
\includegraphics[width=7.7cm,clip]{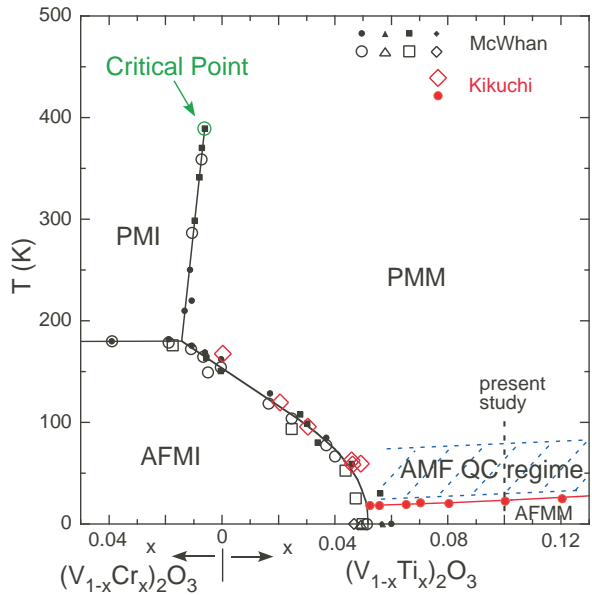}
\end{center}
\caption{
\label{fig:PhaseDiagram}
(color online) 
Phase diagram of (V$_{1-x}$M$_{x}$)$_2$O$_3$ (M = Cr, Ti) 
is reproduced using data points of Refs.~\cite{McWhan73,Kikuchi02}. 
PMI, PMM, AFMI and AFMM stand for paramagnetic insulator, 
paramagnetic metal, antiferromagnetic insulator and 
antiferromagnetic metal phases, respectively. 
AFM QC regime, 
inferred from the present study for $x=0.1$, 
is the temperature range 
where the quantum critical AFM 
fluctuations are controlled by the SDW QPT. 
The AFM transition between PMM and AFMM 
is a second-order phase transition which is tuned 
to the QPT by hydrostatic pressure \cite{Bao98,Carter91}.
}
\end{figure}

The isomorphous weak AFM metals V$_{2-y}$O$_{3}$ \cite{Bao98} 
and (V$_{1-x}$Ti$_{x}$)$_2$O$_3$ \cite{Kikuchi02} 
belong to the celebrated Mott-Hubbard system 
(V$_{1-x}$M$_{x}$)$_2$O$_3$ (M = Cr, Ti) \cite{McWhan73}, which 
shows metal-insulator transitions due to strong correlation effects 
(Fig.~\ref{fig:PhaseDiagram}) \cite{Imada98}. 
The 3$d^2$ electronic state of the V$^{3+}$ ion 
is in an $S=1$ high spin state with 
an effective moment $\sim \! \! 2.8 \mu_{\text{B}}$ \cite{Held01,Mo03}. 
For the AFM metallic (V$_{1-x}$Ti$_{x}$)$_2$O$_3$ ($x > 0.05$), 
only a small fraction of the moment $ \sim \! \! 0.3 \mu_{\text{B}}$ 
forms the AFM ordering below $T_{\text{N}}=23$ K ($x=0.1$) 
\cite{Kikuchi02}. 
The 
second-order 
AFM transition is tuned to a QPT by hydrostatic pressure 
of the order of 2 GPa \cite{Bao98,Carter91}, 
and quantum critical behavior can be expected to be observed 
in the paramagnetic metallic phase. 

Previous neutron-scattering experiments on 
V$_{2-y}$O$_{3}$ clarified several 
interesting aspects of this system \cite{Bao98}. 
At the same time their results raised some 
controversy \cite{SachdevBook99}. 
In those experiments AFM spin fluctuations were roughly 
consistent with a SDW QPT, 
while the data 
suggested 
the $E/T$ scaling indicating a novel QPT. 
However the statistical accuracy of those 
experiments was not sufficient for 
drawing a definite conclusion on the QPT.
Thus in this work we reinvestigate the AFM quantum-critical 
behavior in the paramagnetic metallic phase using 
(V$_{0.9}$Ti$_{0.1}$)$_2$O$_3$ \cite{Kikuchi02}, 
which is suited for the present purpose because 
its local disorder is weaker than in V$_{2-y}$O$_{3}$. 
By sufficiently improving the statistical accuracy, 
we have concluded that the AFM spin fluctuations agree well with 
those of the SDW QPT in three space dimensions. 

Neutron-scattering measurements were performed on the triple-axis 
spectrometers ISSP-GPTAS at the Japan Atomic Energy Agency, 
BT-7 at the NIST Center for Neutron Research, and HB1 
at Oak Ridge National Laboratory (ORNL). 
They were operated using a final energy of $E_{\text{f}}=14$ meV, 
providing an energy resolution of 
1.4 meV (full width at half maximum) at elastic positions. 
A single-crystal sample of (V$_{0.9}$Ti$_{0.1}$)$_2$O$_3$ 
with a weight of 2 g was grown by the 
floating zone method. 
The crystal was mounted in closed-cycle He-gas refrigerators 
so as to measure a $(H,0,L)=H\bm{a}^{\ast}+L\bm{c}^{\ast}$ scattering plane, 
where $\bm{a}^{\ast}$ and $\bm{c}^{\ast}$ are the hexagonal 
reciprocal lattice vectors. 
All the data shown are converted to the dynamical susceptibility 
and corrected for the magnetic form factor.  

\begin{figure}
\begin{center}
\includegraphics[width=5.5cm,clip]{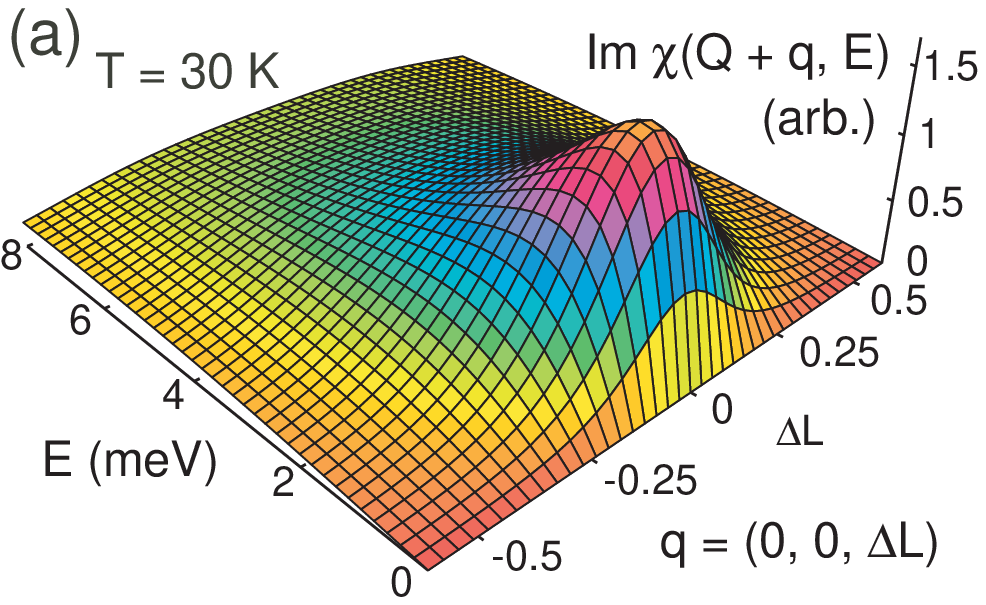}
\includegraphics[width=5.5cm,clip]{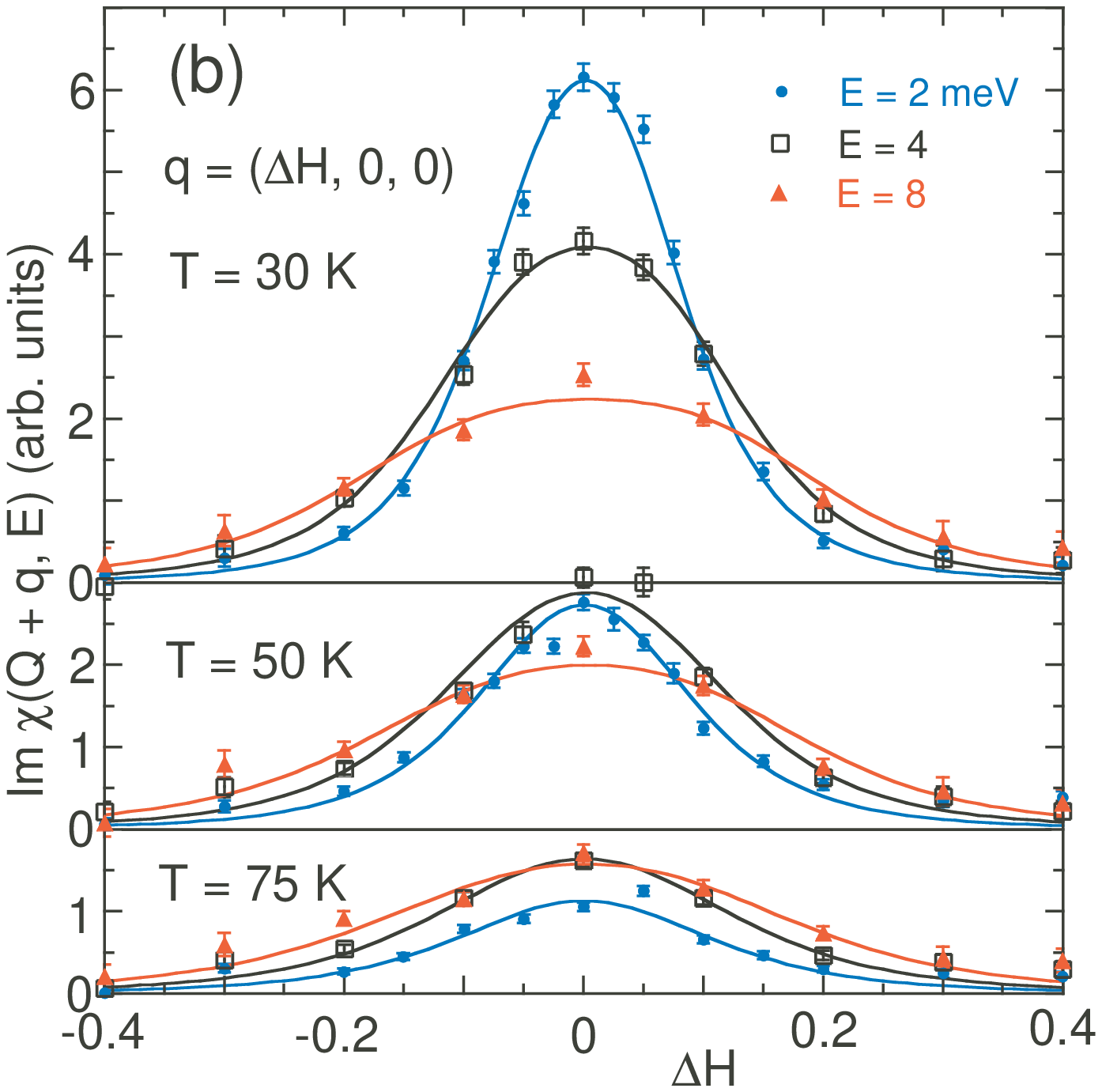}
\includegraphics[width=5.3cm,clip]{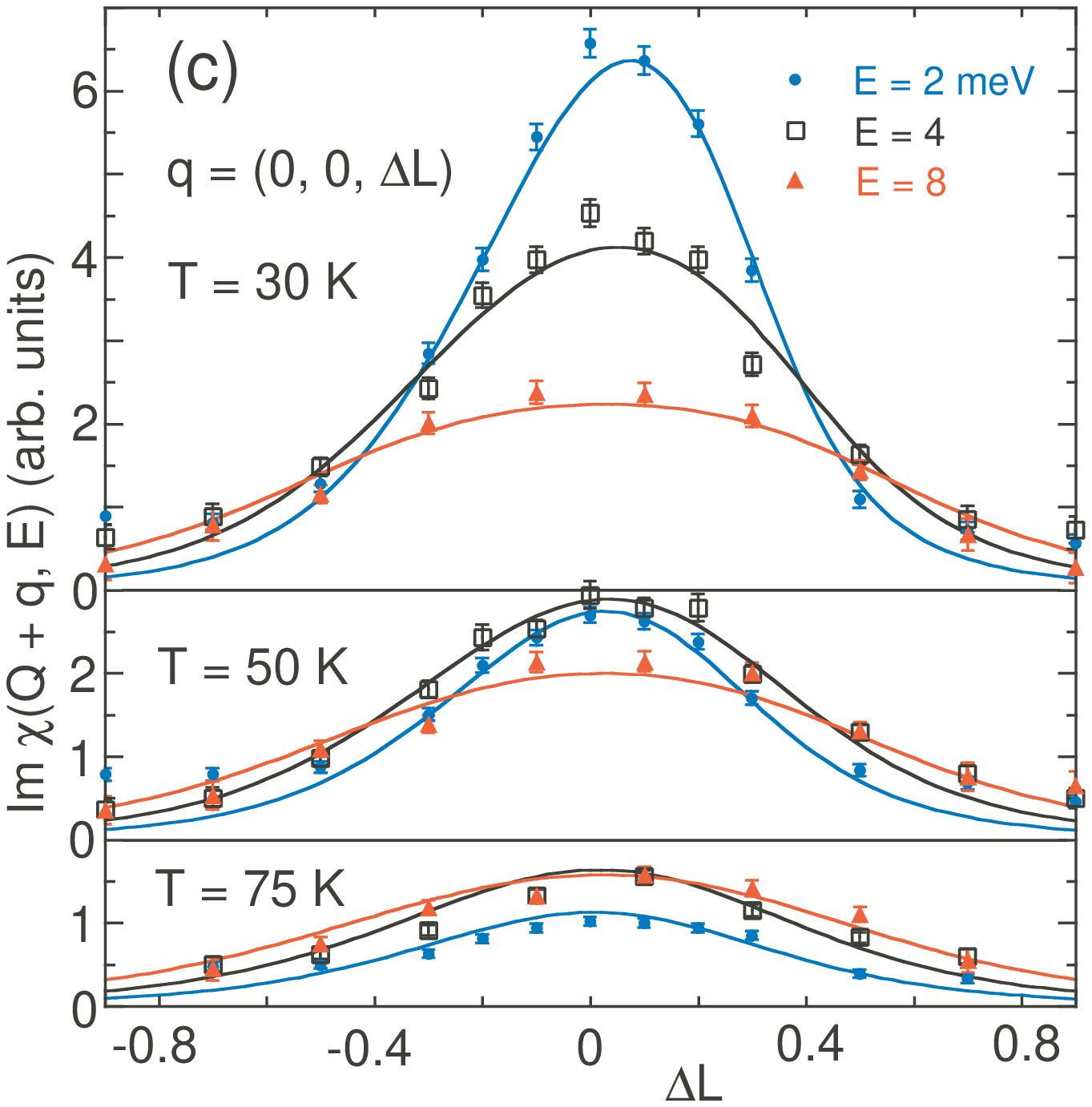}
\end{center}
\caption{
\label{fig:Qscan}
(color online) 
(a) Illustration of the quantum-critical behaviour of 
the dynamical susceptibility 
$\text{Im} \chi (\bm{Q}+\bm{q}, E)$, Eq.~(\ref{eq:ImChiQE}), at $T = 30$ K. 
(b), (c) Constant-$E$ scans taken with $E = 2$, 4, and 8 meV along 
(b) $\bm{q} = (\Delta H, 0, 0)$ and 
(c) $(0, 0, \Delta L)$ lines 
at $T = 30$, 50 and 75 K. 
Curves are fits using Eq.~(\ref{eq:ImChiQE}) 
convoluted with the resolution function. 
}
\end{figure}

The AFM fluctuations of (V$_{0.9}$Ti$_{0.1}$)$_2$O$_3$ 
expressed as the imaginary part of the dynamical susceptibility 
at wave vector $\bm{Q}+\bm{q}$, 
where $\bm{Q} = 
(1.90 \pm 0.01) 
\bm{c}^{\ast}$ is the 
AFM modulation wave vector \cite{Kikuchi02},
are described by the Lorentzian function \cite{Bao98} 
\begin{equation}
\label{eq:ImChiQE}
\text{Im} \chi(\bm{Q}+\bm{q}, E) 
=
\frac{ \chi(\bm{Q}) \Gamma(\bm{Q}) E }
{ E^2 + [\Gamma(\bm{Q}) + D (q_c^2 + F \bm{q}_{ab}^2 ) ]^2 }
\; ,
\end{equation}
where $E$ represents the excitation energy, 
$q_c$ and $\bm{q}_{ab}$ are components of $\bm{q}$
along the $c$ axis and in the $ab$ plane, respectively, 
$D$ and $F$ are $T$ independent parameters, 
$\chi(\bm{Q})$ and $\Gamma(\bm{Q})$ stand for 
the wave-vector-dependent magnetic susceptibility and 
characteristic energy, respectively. 
This form agrees with the approximation used in the theory 
\cite{MoriyaBook85,SachdevBook99,Moriya-Ueda03,Lohneysen07} of 
the SDW QPT for small $\bm{q}$ and $E$, provided that 
the product $ \chi(\bm{Q}) \Gamma(\bm{Q}) $ is $T$ independent. 
We note that $\Gamma(\bm{Q})$ vanishes at a QPT.
In Fig.~\ref{fig:Qscan}(a) the dynamical susceptibility 
Eq.~(\ref{eq:ImChiQE}) is illustrated 
using parameters at $T=30$ K, 
where $\Gamma(\bm{Q})=0.95$ meV. 
To confirm this Lorentzian function for (V$_{0.9}$Ti$_{0.1}$)$_2$O$_3$, 
we carried out constant-$E$ scans along the 
$\bm{q} = (\Delta H,0,0)$ and $(0,0,\Delta L)$ lines 
at three typical temperatures $T = 30$, 50 and 75 K. 
By least squares fitting, we obtained 
$D = 96 \pm 4$ meV \AA$^2$ and $F = 0.77 \pm 0.03$. 
In Fig.~\ref{fig:Qscan}(b) and \ref{fig:Qscan}(c), 
we show these spectra 
together with the fit curves of Eq.~(\ref{eq:ImChiQE}) 
convoluted with the resolution function. 
One can see from this figure that Eq.~(\ref{eq:ImChiQE}) 
well reproduces the experimental data, in particular, for 
small $\bm{q}$ and $E$. 
By this reproduction, we confirmed another assumption 
of Eq.~(\ref{eq:ImChiQE}) that 
$ \chi(\bm{Q}+\bm{q}) \Gamma(\bm{Q}+\bm{q}) $ 
does not depend on $\bm{q}$. 

The theory of the SDW QPT in three dimensions predicts 
\cite{SachdevBook99,Lohneysen07} 
that the characteristic energy $\Gamma(\bm{Q})$ depends on $T$ as
\begin{equation}
\label{eq:T1.5}
\Gamma(\bm{Q}) = c_1 + c_2 T^{3/2} 
\; ,
\end{equation}
where $c_{1} (<0)$ and $c_{2}$ are constants, 
in the quantum critical regime 
$T_{\text{N}} < T \ll T_{\text{coh}}$, where 
the coherence temperature $T_{\text{coh}} \sim $ 450 K \cite{Baldassarre07} 
represents the effective Fermi energy. 
It should be noted that the $T$ dependence of 
$T^{3/2}$ \cite{MoriyaBook85,Hasegawa74} in 
Eq.~(\ref{eq:T1.5}) is the most important 
characteristic of the SDW QPT. 
We also note that Eq.~(\ref{eq:T1.5}) breaks down near $T_{\text{N}}$, 
because the theory neglects the criticality of 
the finite-temperature phase transition. 
In an alternative formalism using 
the self-consistent renormalization (SCR) 
theory of spin fluctuations \cite{Moriya-Ueda03,MoriyaBook85}, 
equivalent to the SDW QPT, 
the $T$ dependence of $\Gamma(\bm{Q})$
is determined by the self-consistent equation 
\begin{align}
\label{eq:SCR}
\Gamma&(\bm{Q}) = 
c_1^{\prime} \nonumber \\
&+ 
F_{Q} 
\int_{0}^{\infty}dE \frac{1}{e^{E/k_{\text{B}}T}-1} 
\sum_{\bm{q}} \text{Im} \chi (\bm{Q}+\bm{q}, E) 
\; ,
\end{align}
where $c_1^{\prime}(<0)$ is a constant and $F_{Q}$ is the mode-mode 
coupling constant. 
This equation employed with Eq.~(\ref{eq:ImChiQE}) and 
$\chi(\bm{Q}) \Gamma(\bm{Q}) =$ const 
can be used as an experimental fit formula, 
where $c_1^{\prime}$ and $F_{Q}$ are 
treated as adjustable parameters. 

\begin{figure}
\begin{center}
\includegraphics[width=7.4cm,clip]{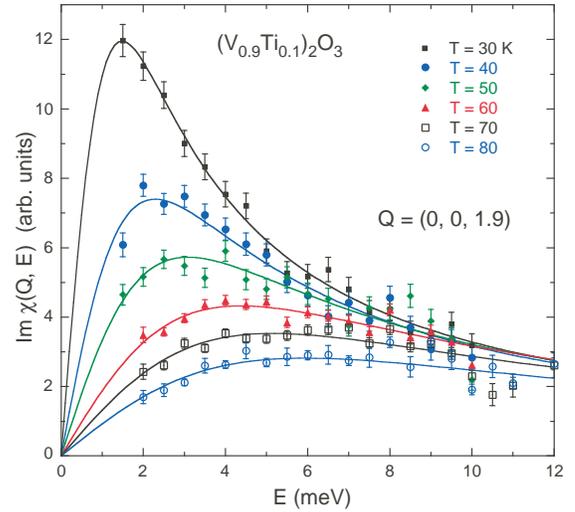}
\end{center}
\caption{
\label{fig:Escan}
(color online) 
Constant-$Q$ scans 
measured at the AFM wave vector 
$\bm{Q} = 1.9 \bm{c}^{\ast}$ at several temperatures. 
Curves are fit results using Eq.~(\ref{eq:ImChiQE}) with 
two adjustable parameters 
$\Gamma (\bm{Q})$ and $\chi (\bm{Q})$. 
Error bars are statistical in origin and represent 1 standard deviation. 
}
\end{figure}

\begin{figure}
\begin{center}
\includegraphics[width=7.8cm,clip]{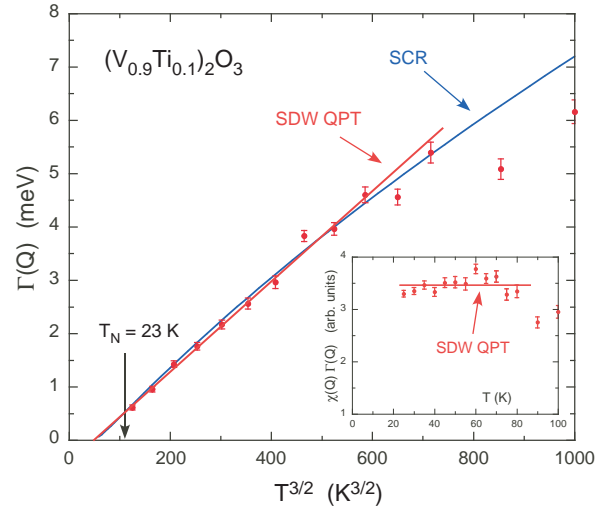}
\end{center}
\caption{
\label{fig:GammaQ}
(color online) 
Temperature dependence of the characteristic energy 
$\Gamma (\bm{Q})$ of the AFM spin fluctuations 
is plotted as a function of $T^{3/2}$. 
The curves represent the prediction Eq.~(\ref{eq:T1.5}) 
for the SDW QPT and the fit using the SCR theory Eq.~(\ref{eq:SCR}). 
The inset shows temperature dependence of 
the product $\chi(\bm{Q}) \Gamma(\bm{Q})$. 
The straight line $\chi(\bm{Q}) \Gamma(\bm{Q}) =$ const is 
the prediction of the SDW QPT.
}
\end{figure}

In order to accurately measure the $T$ dependence of $\Gamma(\bm{Q})$, 
we performed constant-$Q$ scans at the AFM wave vector 
using better counting statistics than Ref. \cite{Bao98}.
The observed spectra were fit to Eq.~(\ref{eq:ImChiQE}) 
convoluted with the resolution function. 
Several spectra and fit curves are shown in Fig.~\ref{fig:Escan}, 
demonstrating excellent agreement between the observation and calculation. 
Figure \ref{fig:GammaQ} shows the $T$ dependence of 
$\Gamma(\bm{Q})$ and 
$\chi(\bm{Q}) \Gamma(\bm{Q})$ 
as a function of $T^{3/2}$ and $T$, respectively. 
The predictions of the SDW QPT, Eq.~(\ref{eq:T1.5}) and 
$\chi(\bm{Q}) \Gamma(\bm{Q}) =$ const, which are also 
plotted using lines in the figure, 
are in good agreement with the experimental data in the 
range $1.1 T_{\text{N}} < T < 80$ K. 
By least squares fitting, we obtained 
$c_1 = -0.37 \pm 0.05 $ meV and $c_2 = 0.0083 \pm 0.0002$ K$^{-3/2}$. 
We also performed the SCR fit using Eq.~(\ref{eq:SCR}), 
where $c_1^{\prime} = -1.1 \pm 0.2$ meV provided the best fit. 
This fit curve shown in Fig.~\ref{fig:GammaQ} also well reproduces 
the experimental data in the same temperature range. 
Therefore, we conclude that the AFM spin fluctuations of 
(V$_{0.9}$Ti$_{0.1}$)$_2$O$_3$ in $1.1 T_{\text{N}} < T < 80$ K, 
which can be regarded as the quantum-critical regime, 
are well accounted for by the quantum-critical behavior of 
the SDW QPT in three dimensions.

It should be noted that the theories of SDW QPTs are based 
upon the single-band Hubbard model in a weak 
correlation regime 
\cite{MoriyaBook85,Hasegawa74,Hertz76,Millis93,Moriya-Ueda03}. 
However the electronic state of V$_2$O$_3$ is 
represented by a three-band model 
with strong correlation \cite{Held01}. 
The two 3$d$ electrons in the V$^{3+}$ ion 
occupying three degenerate $t_{2g}$ orbitals are coupled 
by a strong Hund's rule exchange interaction, which gives rise to 
the $S=1$ state and the orbital degrees of freedom \cite{Held01}. 
The prominent quasiparticle peak at the Fermi energy 
observed by photoemission spectroscopy \cite{Mo03} 
and the low coherence temperature $T_{\text{coh}} \sim 450$ K 
\cite{Baldassarre07} underline the importance of the 
strong correlation in V$_2$O$_3$. 
Thus, the present result poses a natural question 
whether the paramagnetic metallic state of the realistic 
three-band model shows the same quantum criticality 
as the SDW QPT. 
We note that $T_{\text{coh}}$ is comparable to 
the temperature scale $T_0 \sim 320$ K of the 
SCR theory \cite{MoriyaBook85,Moriya94}, and that  
the upper bound temperature 80 K of the quantum-critical regime 
in (V$_{0.9}$Ti$_{0.1}$)$_2$O$_3$ 
may be partly ascribed to orbital fluctuations \cite{Bao98}, 
which are neglected in the theories of the SDW QPT. 

It is widely accepted that 
the correct understanding of AFM QPTs is 
essential for studying unconventional superconductivity 
which has been found in an increasing number of strongly 
correlated electron systems, 
including high-$T_{\text{c}}$ cuprates, 
heavy-fermion and organic superconductors, e.g., 
La$_{2-x}$Sr$_{x}$CuO$_{4}$ \cite{Aeppli97}. 
In these systems, attractive electron couplings were proposed 
to be ascribed to 
AFM spin fluctuations \cite{Moriya-Ueda03,Coleman07}. 
In this context, the spin fluctuations observed in 
(V$_{0.9}$Ti$_{0.1}$)$_2$O$_3$ 
can be considered as a simple nonsuperconducting case \cite{Moriya94}. 

In conclusion, neutron scattering shows that 
the quantum-critical spin fluctuations in 
the paramagnetic metallic phase of the 
Mott-Hubbard system (V$_{0.9}$Ti$_{0.1}$)$_2$O$_3$ 
agree well with the theoretical predictions 
of the SDW QPT in three dimensions. 
The present work is the first clear verification of the SDW QPT 
in a $d$-electron itinerant antiferromagnet. 
The present finding and our recent similar result of an $f$-electron 
heavy fermion \cite{Kado06} imply that a broader theoretical 
basis for the SDW QPT is required to include multiband models and 
strong correlation effects. 
Further investigations of the AFM long-range ordered state close 
to the SDW QPT and crossover phenomena of the QPT to 
finite-temperature phase transitions 
will be interesting.

We acknowledge discussions with T. Moriya and Y. Tabata. 
Work on BT7 and HB1 was supported by the 
US-Japan Cooperative Program on Neutron Scattering. 
The authors are grateful for the local support staff at 
NIST and ORNL. 
The work at ORNL's High Flux Isotope Reactor was sponsored by 
the Scientific User Facilities Division, office of Basic 
Energy Sciences, US Department of Energy.

\end{document}